\documentclass[notitlepage,aps,pra,onecolumn,10pt,superscriptaddress,groupaddress]{revtex4-1}

\usepackage[left=2cm,top=2cm,right=2cm,headsep=0.5cm,headheight=0.5cm,nofoot]{geometry}

\usepackage{mathrsfs,mathpazo,avant,tabularx}
\usepackage[utf8]{inputenc}
\usepackage[table]{xcolor}
\usepackage{colortbl}
\usepackage{indentfirst}
\usepackage{stmaryrd}
\usepackage{amssymb,amsmath,amsthm,amsfonts,amsbsy}
\usepackage{bm,bibunits,color,chngcntr,epsfig,epstopdf,graphicx,dsfont}
\usepackage{hyperref,lipsum,,makecell,mathrsfs,rotating}
\usepackage[english]{babel}
\usepackage[normalem]{ulem}

\newcommand{\be}{\begin{equation}}
\newcommand{\ee}{\end{equation}}
\newcommand{\bea}{\begin{eqnarray}}
\newcommand{\eea}{\end{eqnarray}}
\newcommand{\ket}{\rangle}
\newcommand{\bra}{\langle}

\newcommand{\I}{\mathds{1}}
\newcommand{\ra}{\rightarrow}

\def\C#1{\mathcal #1}

\def\S#1{\mathscr #1}

\definecolor{gray}{gray}{0.9}

\begin{document}
\newtheorem{theorem}{Theorem}
\newtheorem{prop}[theorem]{Proposition}
\newtheorem{corollary}[theorem]{Corollary}
\newtheorem{open problem}[theorem]{Open Problem}
\newtheorem{conjecture}[theorem]{Conjecture}
\newtheorem{definition}{Definition}
\newtheorem{remark}{Remark}
\newtheorem{example}{Example}
\newtheorem{task}{Task}

\title{A survey of universal quantum von Neumann architecture}

\author{Yuan-Ting Liu} 
\affiliation{CAS Key Laboratory of Theoretical Physics, Institute of Theoretical Physics, \\
Chinese Academy of Sciences, Beijing 100190, China}
\affiliation{School of Physical Sciences, University of 
Chinese Academy of Sciences, Beijing 100049, China}
\author{Kai Wang} 
\affiliation{CAS Key Laboratory of Theoretical Physics, Institute of Theoretical Physics, \\
Chinese Academy of Sciences, Beijing 100190, China}
\affiliation{School of Physical Sciences, University of 
Chinese Academy of Sciences, Beijing 100049, China}
\author{Yuan-Dong Liu} 
\affiliation{CAS Key Laboratory of Theoretical Physics, Institute of Theoretical Physics, \\
Chinese Academy of Sciences, Beijing 100190, China}
\affiliation{School of Physical Sciences, University of 
Chinese Academy of Sciences, Beijing 100049, China}
\author{Dong-Sheng Wang \footnote{Corresponding author. Email: wds@itp.ac.cn} }
\affiliation{CAS Key Laboratory of Theoretical Physics, Institute of Theoretical Physics, \\
Chinese Academy of Sciences, Beijing 100190, China}

\begin{abstract}
The existence of universal quantum computers has been theoretically 
well established. 
However, building up a real quantum computer system not only 
relies on the theory of universality, 
but also needs methods to satisfy requirements on 
other features, such as programmability, modularity, scalability, etc. 
To this end, we study the recently proposed model 
of quantum von Neumann architecture, 
by putting it in a practical and broader setting, 
namely, the hierarchical design of a computer system. 
We analyze the structures of 
quantum CPU and quantum control unit, 
and draw their connections with computational advantages. 
We also point out that a recent demonstration 
of our model would require less than 20 qubits. 
\end{abstract}

\maketitle


\section{Introduction}


At the origin of quantum computing, physicists such as R. Feynman and D. Deustch 
realized that universal quantum computing is possible~\cite{Fey82}.
We shall also notice that at that time classical computers were just being built. 
After decades of evolution, 
classical computers have become more and more advanced. 
At the meantime, the field of quantum information science grows,
and nowadays physicists and engineers can control quantum processors 
of tens or even hundreds of qubits. 

As the foundation of computation, 
physics is not only crucial to guide the finding of elementary devices
such as transistors, 
but also crucial to set the principles of computation  
regarding space, time, energy, efficiency, etc. 
However, physics itself is not enough.
For the building of classical computers, 
some other disciplines of study also played essential roles,
in particular, 
the theories of information, system, and control.
The information theory, established by C. Shannon~\cite{Sha48}, borrows ideas from thermodynamics 
but it reveals far more properties of information.
The system theory, pioneered by von Bertalanffy~\cite{Ber72}, has more connections with many-body physics
and it emphasizes more on the structure, correlation, etc 
rather than the individual participant.
The control theory, with N. Wiener~\cite{Wie48}, studies the interplay 
between the controller and the target system to achieve a certain goal. 
These studies go beyond the traditional scope of physics.
Together with computational complexity theory~\cite{AB09book},
they form the theoretical foundation to make classical computers real.

The modern quantum physics, especially quantum information science,
is not a traditional physics; instead, 
it shares features of engineering.
It does not only study a system passively,
namely, only study those that exist naturally, 
but also actively study a system, 
e.g., how to make an artificial system for a certain purpose. 
Therefore,
it needs both physicists and engineers to make 
quantum computers real, too. 

The power of quantum computing is currently mainly demonstrated by quantum algorithms.
Given a problem, 
a quantum algorithm is constructed in the framework of a universal quantum computing model,
such as the circuit model~\cite{Deu85}, or a Hamiltonian-based model~\cite{Fey82,Llo96}.
An algorithm is realized by a sequence of elementary operations available in a model,
such as the CNOT gates and qubit gates~\cite{NC00}. 
However, from a modern design of computers~\cite{HH13},
the above is not enough to guide the design of a real quantum computer. 
A computer system is far more complicated than a physical experimental device. 
From the hierarchy of the layers of abstraction,
the physical devices and gates are at the bottom layer of the hierarchy,
while algorithms and applications are at the top layer of it.
See Fig.~\ref{fig:layer}.
There is a gap between them. 
Some investigations on quantum high-level programming and layered design have been 
taken; e.g., Refs.~\cite{SAC+06,HSS+18}.
Filling up the gap, although may take decades,
is necessary to build real quantum computing systems.

To this goal, we need to understand how to satisfy the requirements of 
programmability, modularity, automation, etc besides the basic requirement of universality. 
A computing device or system is programmable 
if it can realize a broad range of programs (or algorithms) 
without almost any change of its physical structure. 
That is, programs can be loaded as software.
A system is modular if the connections among different parts of it, known as units, 
are device-independent,
namely, a unit can be detached or replaced without affecting other units. 
A system is automate if it can realize hierarchical or concatenated tasks  
without active interfering in the middle. 
Realize these features have greatly benefited modern computers and also engineering. 

With the methodology above, 
in this work we present a survey of quantum von Neumann architecture.
There were explorations on this subject in literature~\cite{MWY+11,Bra17,KSM+20},
however, they did not present a universal model with explicit stored quantum programs. 
Based on channel-state duality~\cite{Cho75},
a universal model for quantum von Neumann architecture
is recently developed~\cite{W20_choi,W22_qvn,W23_ur,W23_qvn}.
In this work, we further study it by focusing on a few subjects, 
especially the structure of the quantum CPU, also known as QPU,
and the structure of the quantum control unit (QCU). 
We also survey the elementary requirements for a near-term
demonstration of this architecture. 
Our study is purely theoretical
without referring to any actual quantum computing platforms. 
With this survey, we hope to explain some details of the model
and identify some research directions to investigate in the near future. 

This work contains the following parts.
In Section~\ref{sec:cc},
we first review the principle of classical computer.
We then review the basics of quantum computing in Section~\ref{sec:qc}. 
We then survey the basic operations in quantum von Neumann architecture (QvN) in Section~\ref{sec:qvn}.
After these, in Sections~\ref{sec:diffq}
we discuss the features of our model compared with other quantum computing models.
In Section~\ref{sec:alg},
we survey algorithm designs in QvN and their possible computational advantages.
We then study the basic requirement for a NISQ implementation of QvN in Section~\ref{sec:nisq}.
We then conclude in Section~\ref{sec:conc} with open questions and perspectives.

\section{Classical computer}\label{sec:cc}

It would be interesting to review how a classical computer is built, mainly 
what are the underlying principles. 
This will help to understand what are the current situations for quantum computing. 
In this section, we start from a few universal computing models, 
and then move on to the layers of structures for the design of a modern computer.

\subsection{Computing models}\label{subsec:model}

A universal computing model is a framework to design algorithms for solving problems. 
The most well known model is the circuit model based on Boolean logic,
while at the same time 
there are a few equivalent ones, including the Turing machine, 
cellular automata, etc. 
Their logic building blocks are different, but can simulate each other efficiently~\cite{AB09book}. 

We start from the circuit model. 
Information or data are represented as bit strings,
and the basic Boolean gates on bits include NOT, AND, NAND, OR, etc. 
An important theorem is that there exist a universal set of gates 
so that any Boolean function $f: \{0,1\}^n\mapsto \{0,1\}$ can be expressed as 
a sequence of these gates, forming a circuit. 
Such circuits are not invertible as some bits are lost, 
but they can be made invertible by using the Toffoli gate to simulate them.
The Toffoli gate is 
\be \text{Tof}= P_0 \otimes \I + P_1 \otimes \text{CNOT}, \ee 
for the controlled-not gate 
\be \text{CNOT}= P_0 \otimes \I + P_1 \otimes \text{NOT}, \ee
with $P_0$ ($P_1$) as projection on bit-value 0 (1).
Despite this, a circuit is often designed using the Boolean gates. 

The circuit model is fundamental for the characterization of universality 
and also the design of algorithms. 
However, it is still abstract without specifying components for building a real computer. 
The foundation for the design of modern computers is the so-called 
von Neumann architecture (vNA)~\cite{Neumann1993FirstDO},
which contains a few modular parts, 
including the central processing unit (CPU), memory, control, internet, and in/out units. 
All these can be described by the circuit model,
but it is crucial to separate them.
In particular, the stored programs in the memory unit 
are essential to realize universality and programmability.
Namely, a stored program as bit strings can be read and then loaded to 
the programmable CPU, without physically changing the structure of the CPU
in order to run different algorithms (i.e., programs). 
Formally, this realizes 
\be G (\vec{b} \times \vec{b}_A ) = A \vec{b} \times \vec{b}_A',\ee
for $G$ as the CPU, $\vec{b}$ as an input bit string,
$\vec{b}_A$ as the bit-string encoding of an algorithm $A$.
The desired output is $A \vec{b}$.
The final $\vec{b}_A'$ is often ignored but can be used to recover $\vec{b}_A$. 
The program also contains control signals for precise timing and addressing
of data and operations or commands. 
Although the internet was invented later than the vNA itself,
and there are also many types of communication,
the download and upload of data is an indispensable part of vNA. 

Although it seems vNA is a step closer to a real computer than the circuit model itself,
vNA is still an abstract model. 
A modern computer is far more complicated than the abstraction of vNA.
In particular, it follows a hierarchical design of layers of abstraction,
with the physical logic devices at the bottom, and algorithms and applications at the top. 
For instance, 
there are many types of memory, such as the internal storage, 
hard disk (as external storage), and flash memory etc
playing distinct roles in computers and also microchips. 

\subsection{Hierarchical design}

\begin{figure}
    \centering
    \includegraphics[width=0.4\textwidth]{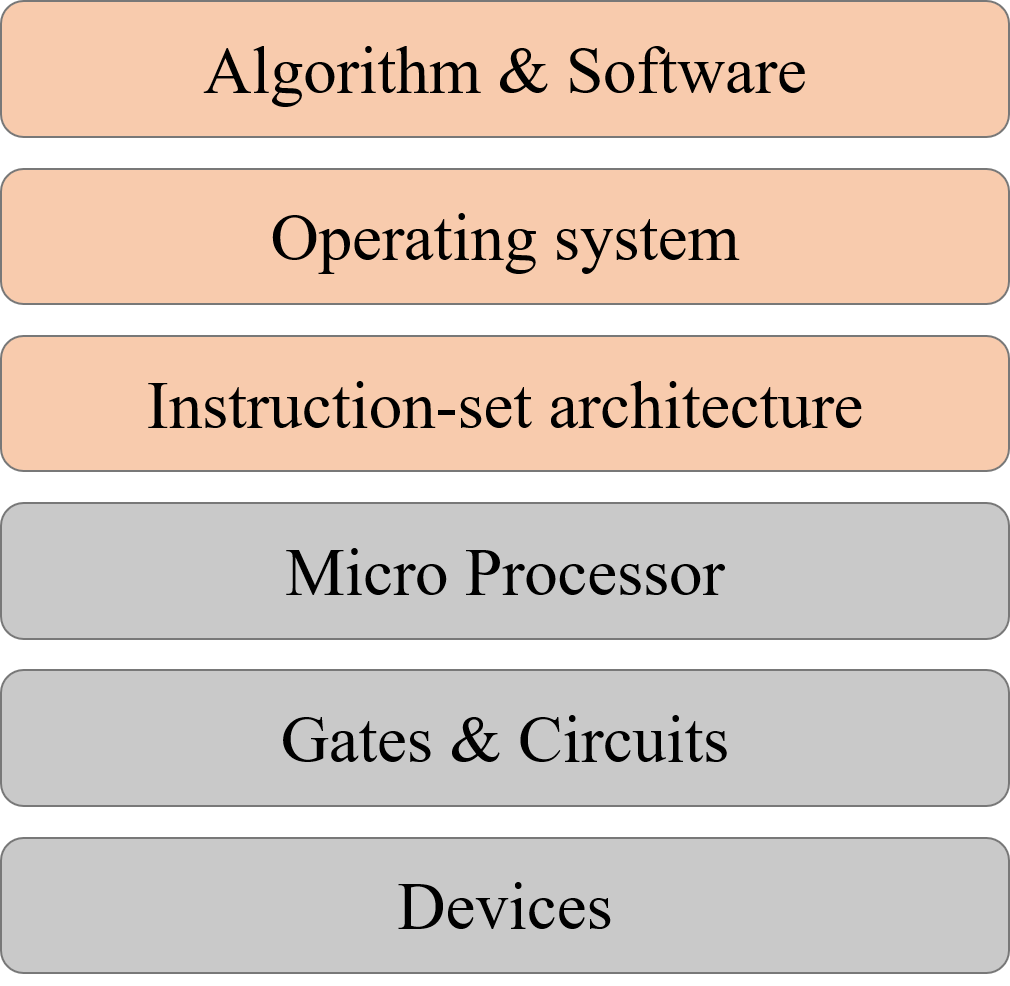}
    \caption{Layers of abstraction of a computer.}
    \label{fig:layer}
\end{figure}

The hierarchical layers of abstraction for a computer architecture 
is a crucial step to build a real universal programmable computer~\cite{HH13}.
Here we take a brief overview of it, mainly the physical aspects, 
also see the Figure~\ref{fig:layer}. 
It contains both hardware and software layers, 
with different programming languages associated with them. 

\begin{itemize}
    \item The physical devices: this is to find physical system as the carrier of bits. 
    For instance, they are the transistors as the basic element to construct logic gates, 
    or the magnetic domain for storage. 
    \item The gates and circuits: this is to design the elementary Boolean gates and also elementary circuits,
    such as the adder, multiplexer, decoder, and a few sequential circuits, 
    such as Latch, Flip-Flop, Register. 
    This is on the level of machine language. 
    \item The micro-processor: this is based on vNA.
    It can realize instructions such as if, else, when, while, for, shift, branch, etc for the purpose of programming. 
    \item The instruction-set architecture: this is to design instructions, operand locations, memory, control, etc;
    such as CISC, MIPS. 
    This is on the level of assemble language. 
    \item The operating system: this decides how people can use a computer, 
    such as how to make input/output, how to input command.
    \item The algorithm and software: these are programs for solving certain class of problems. 
    This is on the level of advanced language. 
\end{itemize}

From the hierarchy of the layers of abstraction, 
we can see that quantum computing is still at an early stage. 
Currently, what people mostly aim at is quantum CPU that can run 
simple circuit-level quantum algorithms, 
while all other parts can be classical. 
Namely, it uses classical control, classical memory, and also classical operating system. 
The quality of qubits and also circuits are getting better,
but these are at the lower levels of the hierarchy. 
There is no real logical qubit yet,
which shall be error-correcting, either self-correcting or actively. 
It is still at the infancy to construct quantum micro-processor and instruction-set architecture,
and this requires a better understanding of the roles of quantum memory and quantum control,
and the roles of being quantum in other devices. 



\section{Basics of quantum computing}\label{sec:qc}

In this section, we briefly review the basics of quantum computing~\cite{NC00}
and set the stage for our study. 
We focus on finite-dimensional Hilbert spaces. 
For a Hilbert space $\S H$, 
quantum states are known as density operators $\rho \in \S D(\S H)$,
forming a convex set of nonnegative semi-definite operators with trace 1.
A state is pure if it is also a projector. 
Quantum evolution is 
in general described by completely-positive trace-preserving (CPTP) maps~\cite{Kra83},
or known as quantum channels. 
A fundamental principle is the quantum channel-state duality,
i.e., the Choi-Jamio{\l}kowski isomorphism~\cite{Jam72,Cho75} that
maps a channel $\mathcal{E}$ into a quantum state
\begin{equation}\label{eq:choi}
\omega_\C E:=\C E \otimes \mathds{1} (|\omega\rangle\langle\omega|),
\end{equation}
called Choi state in this work, 
for
\be |\omega\rangle:=\frac{1}{\sqrt{d}}\sum_{i=0}^{d-1}|i,i\rangle  \in \S H \otimes \S H\ee
as a Bell state, known as an ebit, with $d=\text{dim}(\S H)$.

A channel can also be written as a Kraus operator-sum representation
\be \C E(\rho)=\sum_{i=1}^r K_i \rho K_i^\dagger, \ee
for $K_i$ as Kraus operators~\cite{Kra83} with $\sum_i K_i^\dagger K_i=\I$.
This can be found from the eigen-decomposition of $\omega_{\C E}$, 
and $r$ is the rank of $\omega_{\C E}$.

Unitary evolution and quantum measurement can both be viewed as channels.
A unitary $U\in SU(d)$ is rank 1 with $U^\dagger U=UU^\dagger=\I$.
Its dual state is a pure state
$|\omega_U\rangle=( U \otimes \mathds{1}) |\omega\rangle$,
and we will use the notation $|U\ket$ for simplicity. 
A quantum measurement is a POVM (positive operator-valued measure),
which is a set of positive operators $\{M_i\}$ with $\sum_i M_i=\I$.
It is clear to see each effect $M_i$ can be realized as 
$M_i=K_i^\dagger K_i$ for a Kraus operator $K_i$,
therefore the POVM is realized by a channel. 

A channel can be described as an isometry $V$ with $V=\sum_i |i\ket K_i$,
so it can be realized by a unitary $U$ with $V=U|0\ket$ as the first block column of $U$,
and $|0\ket$ as the initial ancillary state. 
This is the Stinespring's dilation theorem,
which guarantees that it is enough to consider unitary evolution on pure states,
since non-unitary channels and mixed states can be realized by ignoring ancilla or subsystems. 

In quantum circuit model, we consider unitary evolution on multi-qubit states followed by measurement.
Similar with the classical circuit model,
there also exist universal gate sets to decompose arbitrary unitary operator~\cite{BBC+95}.
Two well known examples are the set $\{$H,T,CNOT$\}$ and $\{$H,Tof$\}$,
for the Hadamard gate H, T gate as the forth-root of Pauli Z operator.
The Toffoli gate is universal for classical computing, 
but with H gate, they are universal for quantum computing.
The H gate exchanges Pauli X and Z operators 
\be HXH=Z, HZH=X, \ee 
while with $T^2$, which is the phase gate S, and the CNOT,
they form the Clifford group~\cite{GC99} that preserves the set of (tensor-product of) Pauli operators.
It is known that Clifford circuits are not even universal for classical computing.
The non-Clifford gates such as the T and Tof are necessary to achieve quantum universality.

We see that quantum measurement is needed to read out the results,
which can be viewed as expectation value of observable on the final state. 
It is also possible to encode expectation values as bit strings 
and require the final state of quantum algorithms to be bit strings,
such as using the amplitude amplification algorithm~\cite{BHM02},
but this will cost more quantum resources. 
To estimate expectation values, we often run the same circuit multiple times 
to get the necessary probabilities.  
Namely, to measure $\text{tr}(A \rho)$ for a hermitian observable $A$ on the final state $\rho$,
the eigenspectrum of $A=\sum_i a_i |i\ket\bra i|$ is needed, 
and probabilities $p_i=\bra i|\rho|i\ket$ are obtained by repeated measurments so that 
\be \text{tr}(A \rho)= \sum_i p_i a_i. \ee

That is, there are two primary but fundamental differences between the classical and quantum cases:
the quantum evolution is unitary but non-unitary measurement is needed for readout.
It is more appropriate to treat quantum algorithms as extensions of probabilistic algorithms,
which not only use Boolean circuits acting on bits but also random numbers, 
in the form of pbits. 
Qubits can be understood as a combination of bits and pbits in the sense that
its basis for a Hilbert space are bits while its ampitudes in this basis 
are the source of pbits. 

\section{Basics of quantum von Neumann architecture}\label{sec:qvn}

\begin{figure}[t!]
    \centering
    \includegraphics[width=0.45\textwidth]{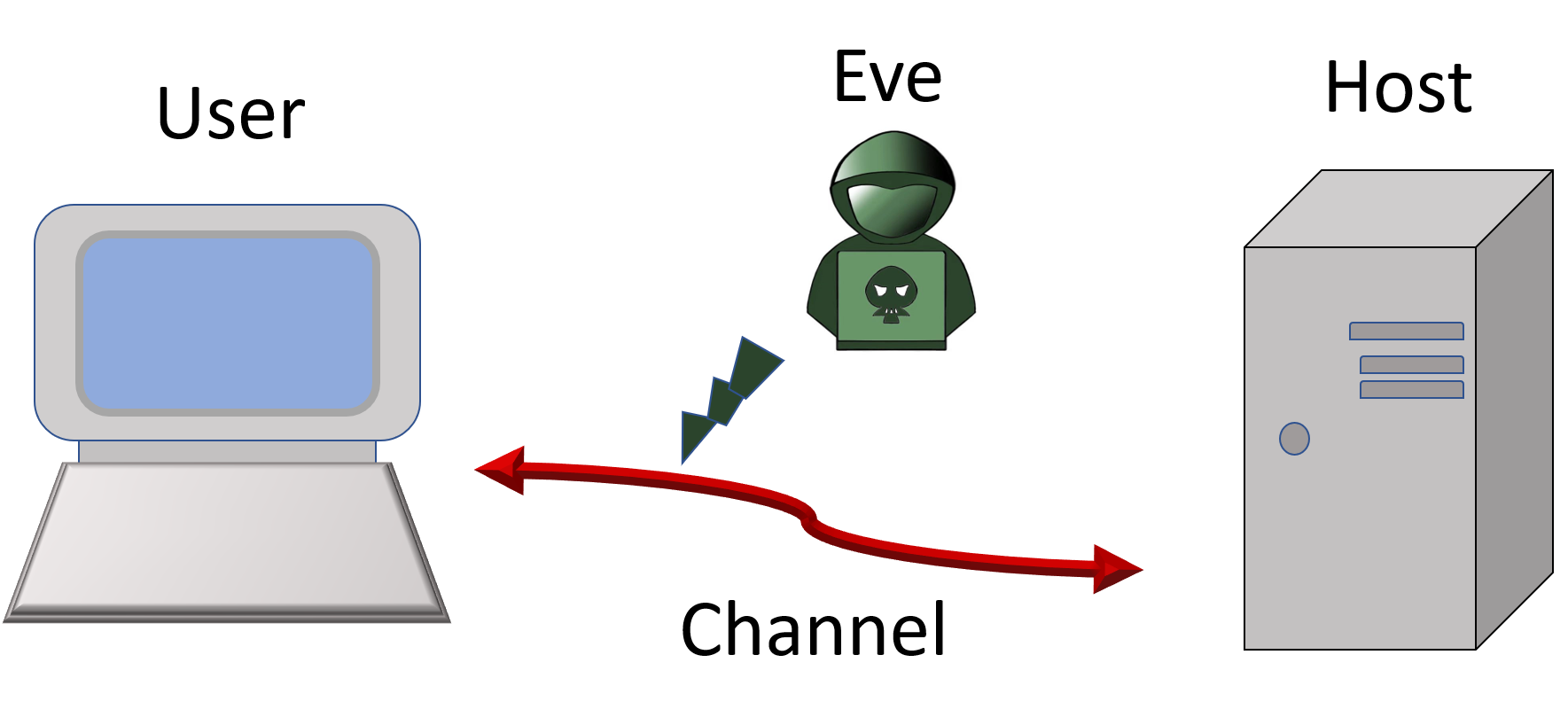}
    \caption{The model we use for quantum von Neumann architecture in this work. 
    The user aims to perform a quantum algorithm, 
    while the algorithm or program is provided by the host through a quantum channel, 
    which can be monitored by an eavesdropper Eve.}
    \label{fig:model}
\end{figure}

In this section, we discuss the basic model of
quantum von Neumann architecture (QvN) based on our recent work~\cite{W20_choi,W22_qvn,W23_ur,W23_qvn},
and here we aim to explain the details of the elementary operations in our model.
Note we do not study how to physically construct or encode a logical qubit,
or physically construct a unit,
which are separate important subjects.

\subsection{The basic model} 


We describe the basic model as shown in Fig.~\ref{fig:model}.
This is the analog of what exists nowadays for modern computer system.
Of course, we only discuss the primary abstract process. 
A user aims to perform a quantum algorithm, 
while the algorithm or program is provided by the host through a quantum channel, 
which can be monitored by an eavesdropper Eve,
or suffers from noises.
Quantum codes are needed to protect information against noises and Eve,
and they are also needed for the computers.

In practice, a host or data centre may have a different design from a desktop computer.
However, for simplicity we assume a quantum host follows a similar design with a quantum computer.
The program may come from a host or another user. 
Without digging into the structures of a user or host computer, 
below we explain the elementary operations that need to be performed. 

\subsection{Read and write on memory}

Given a quantum program encoded in a quantum state, 
one has to execute it.
Using Choi state, the underlying scheme is that 
the action of a channel $\C E$ on state $\rho$ is recovered as
\be \C E(\rho)= d \; \text{tr}_\text{B} [\omega_{\C E} (\I \otimes \rho^t) ] \label{eq:readout}\ee
for $\rho^t$ as the transpose of a state $\rho$.
The partial trace $\text{tr}_\text{B}$ is on the 2nd part of $\omega_{\C E}$.
Below and most of the time in this work, we only consider unitary programs.
A program $U$ is stored as its Choi state $|U\ket=U\otimes \I |\omega\ket$.
See the figure 
\begin{figure}[h!]
    \centering
    \includegraphics[width=0.1\textwidth]{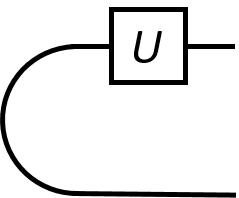}
\end{figure}

\noindent The curve is the Bell state $|\omega\ket$.
Given $|U\ket$, how to use it?
The basic usage is how it acts on input.
In our scheme, an input state is injected by a binary projective measurement,
and the output is obtained also by a projective measurement (PVM).

Suppose the initial state is $|0\ket$, and we need to obtain 
\be p_i=|\bra \psi_i|U|0\ket|^2. \ee
The binary PVM for initial-state injection is $\{P_0, P_{\bar{0}}\}$ 
for $P_{\bar{0}}=\I - P_0$.
The PVM for readout is $\{|\psi_i\ket \bra \psi_i|\}$.
As measurement outcomes are random, the initial state is only realized with finite probability.
However, this is not a problem.
For the case of $P_0$, we obtain $p_i$.
For the case of $P_{\bar{0}}$, we get 
$p_i'=1-p_i$ so $p_i$ can also be obtained~\cite{W20_choi}.

If the dimension of $U$ is $d$, then we need a qubit-ancilla to realize the binary PVM.
See the figure for $n$-qubit input:
\begin{figure}[h!]
    \centering
    \includegraphics[width=0.2\textwidth]{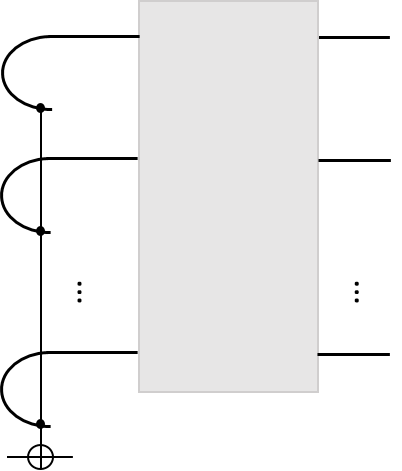}
\end{figure}

\noindent The Toffoli-like gate (on the left) is needed to extract the parity to the ancilla.
A PVM on the ancilla will realize the initial-state injection.
For convenience, we often call the 2nd part of a Choi state as the `tail',
which serves as the `port' for the initial-state injection, a `write' operation,
and the 1st part as the `head', 
which is the port for `read' operation. 

Besides the Choi-state form, there are also other ways to store a program.
Note that a program $U$ 
can be decomposed as a circuit of elementary gates $U\approx \prod_i U_i$
with a fixed accuracy $\epsilon$.
Here we discuss a few of them.

\begin{itemize}
    \item A quantum encoding: use the Choi state $|U\ket$.
    \item A classical encoding: use bits $[U]$ to represent $U$ as a matrix, 
    or as a sequence of gates forming a circuit, with 
    the location and type of each gate encoded by bits.
    \item A hardware encoding: a gate is stored in a hardware device, just like the optical elements in photonic quantum computing~\cite{NC00}.
\end{itemize}

Different schemes can be applied in different settings.
They will also affect the construction of the QPU. 
Note there are also other ways. 
There is a nonlocal Choi-state-like form so that a program 
can be executed blindly~\cite{YRC20},
but this requires a lot more resources,
therefore, we do not study this form in this work. 
The classical encoding $[U]$ is most popular nowadays. 
It can be used as classical control signals 
to guide the execution of gates. 
This applies to the current framework on circuit model,
such as superconducting circuits. 
Below we will study how to use the quantum encoding to construct QPU.


\subsection{Teleportation}\label{subsec:tele}

Teleporation has been used in many ways, e.g.,
in quantum communication,
in fault-tolerant scheme and in measurement-based quantum computing.
For QvN, teleporation is used both for communication and computation. 
In communication, it has been well established that 
teleporation can replace the transmission of qubits by bits 
given distributed ebits~\cite{BBC+93}. 
For computation, teleportation is employed to realize gate operations,
similar with the measurement-based model.
Here we recall its definition and motivate the covariant teleportation. 


One often starts from a bipartite nonlocal setting that
Alice and Bob already shared ebits, and Alice aims to send qubits (or qudits)
to Bob without quantum communication.
The scheme is shown in the figure 
\begin{figure}[h!]
    \centering
    \includegraphics[width=0.3\textwidth]{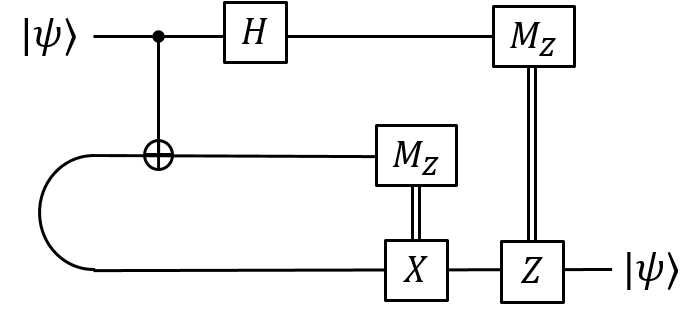}
\end{figure}

The Pauli byproducts $\sigma_i\in \{I, X, Y, Z\}$ are corrected by sending the measurement outcomes $i$
from the Bell measurement of Alice to Bob.

There is an interesting symmetry of this scheme.
Each Pauli byproduct is obtained with the same probability. 
The operators $\sigma_i$ form a projective representation of the group $Z_2\times Z_2$.
Actually, this fact has been used to define group-based teleportation~\cite{BDM+00}.

This can also be understood from the point of view of tensors.
The set of Pauli byproducts form a three-leg tensor, 
and it has the full symmetry $SU(2)$ if the identity operator is absent~\cite{W19_rev}. 
This also applies to any $SU(d)$ and leads to the covariant teleportation~\cite{W20_choi}  
by grouping the nontrivial Pauli byproducts together,
namely, using a qubit ancilla to extract the binary distinction of byproducts.




\subsection{Switchable composition of programs}\label{subsec:comp}


The covariant teleportation, also called universal quantum teleporation (UQT),
can be used to compose two programs together. 
Namely, two program states $|U\ket$ and $|V\ket$ can be composed 
together deterministically to be $|UV\ket$, 
or $|VU\ket$ depending on the direction of information flow.
See the figure: 
\begin{figure}[h!]
    \centering
    \includegraphics[width=0.3\textwidth]{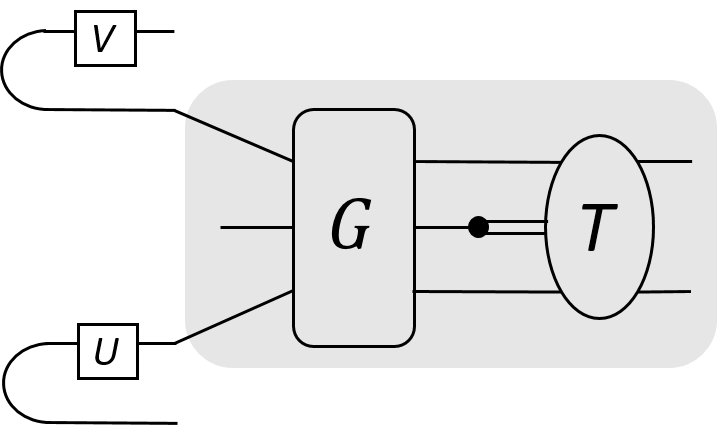}
\end{figure}

The shaded region is for the UQT. 
It requires a qubit ancilla and Toffoli-like gate, 
and also the adjoint form $T$, also known as the affine form, of a gate~\cite{W20_choi}.
For instance, a qubit gate $U\in SU(2)$ corresponds to an orthogonal rotation $R\in SO(3)$.
A PVM on the qubit ancilla leads to either trivial or nontrivial Pauli byproducts,
conditioning on which the correction $T$ is applied.
Note that in order to complete composition, 
the programs need to be known, i.e., as white boxes. 
This can be used to generate large programs from smaller ones.
When only elementary programs are composed, such as $|H\ket$, $|T\ket$, $|CZ\ket$ 
for the Hadamard gate H, T gate, and CZ gate,
only the adjoint form of H and T needs to be done. 
As H exchanges Pauli X and Z, while T can generate superposition of Pauli X and Y, 
it is easy to see the affine form of H is a swap gate, 
while of T is a Hadamard-like gate~\cite{W20_choi}.

The ebits used in the composition have a unique feature.
A state injected at its tail can propagate `backward' to its head,
following from the channel-state duality. 
This leads to a switchable construction of the composition.
For instance, for a qubit program it attaches one ebit to it.
Then it applies a few CZ gates, as shown in the figure  \\

\begin{figure}[h!]
    \centering
    \includegraphics[width=0.25\textwidth]{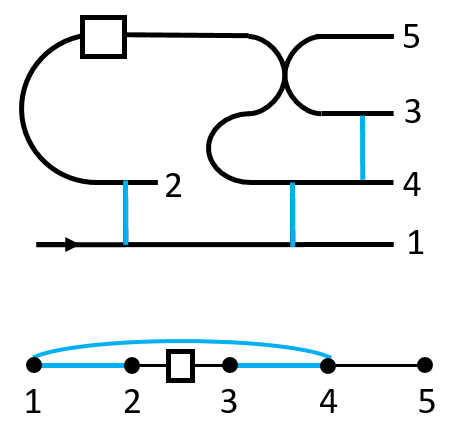}
\end{figure}

\noindent The top panel shows the circuit,
while the bottom panel shows the operations on the qubits explicitly. 
Blue lines are CZ gates. 
The box is the stored program.
This forms a `pre-compose' step between the previous program and the current one.
There are then two possible paths for the information flow, 
one with the current program, 1$\ra$2$\ra$3$\ra$4$\ra$5,
the other without it, 1$\ra$4$\ra$5.
This serves as a switch for turning on or off of the gate
depending on the control signal.
To complete the composition, one path needs to be chosen while closing the other,
and there will also be correctable Pauli byproduct after the composition. 
This will be used to construct the QPU. 

\subsection{Program conversion}\label{subsec:conv}

It is also useful if a program can be changed into another.
This needs the operation of quantum superchannel~\cite{CDP08a,CDP08,CDP09}.
For notation, we use a hat on the symbols for superchannels.
The circuit representation of a superchannel is 
\be \hat{\C S} (\C E)(\rho)= \text{tr}_a \C V \; \C E\; \C U (\rho\otimes |0\ket \bra 0|),  \ee
for $\rho \in \S D(\S H)$, 
$\C U$ and $\C V$ are unitary,
and a is an ancilla. 
The dimension of $V$ can be larger than $U$~\cite{WW23},
but we do not need the details here. 
This can also be represented as the action on Choi state with 
\be \hat{\C S} (\C E)(\rho)=\text{tr}_{\bar{\text{A}}} \C V \otimes \tilde{\C U} 
(\omega_{\C E} \otimes \omega) (\I\otimes \rho^t\otimes |0\ket \bra 0|). \ee
The unitary $\tilde{\C U}$ is the transpose of $\C U$ conjugated by a swap. 
The trace is over the subsystems except the top one, A.
We see that ebits are needed in order to realize nontrivial superchannels.
See the figure 
\begin{figure}[h!]
    \centering
    \includegraphics[width=0.2\textwidth]{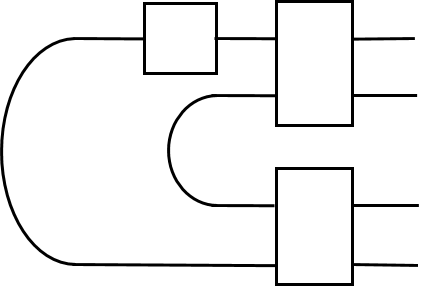}
\end{figure}

\noindent The top wire carries the output.
It has been shown that~\cite{W22_qvn}, a sequence of superchannels acting on Choi states
can be composed together with the tool of UQT. 
This realizes the so-called quantum comb which indeed is a composition of superchannels. 
This has found applications in quantum estimation, learning, optimization etc,
and we will further study this in the Section~\ref{sec:alg}.

\subsection{QCU}\label{subsec:cont}



Control plays essential roles in classical computers.
The simplest example is the CNOT gate, 
which uses a bit to control another bit.
Clock is another notable example which is a building block
in classical sequential circuits.
There are also schemes which use electric circuits 
to achieve control of analog signals. 
Here we analyze the construction of quantum control unit (QCU) for QvN.

First, there are different layers of control tasks. 
The most familiar one is to use bits to control quantum gates.
In the circuit model, each gate has definite spacetime location, i.e.,
when and where it acts on qubits. 
Such classical information serves as bits to control the execution of quantum gates.
There is no entanglement between the control bits and data qubits.

A semi-classical scheme is to use lasers to interact with qubits, 
a seminal field of AMO physics and also the most familiar paradigm of quantum control. 
There is no entanglement between the qubits and the lasers. 
The dynamical decoupling~\cite{VKL99} is a notable example.

One can also use qubits to control quantum gates. 
This actually has been a quite common scheme for designing quantum algorithms,
such as the swap test, DQC1, and also quantum phase estimation~\cite{W23_ur}.
This can lead to interference of quantum gates,
and this has been used in the linear combination of unitaries (LCU) algorithm~\cite{CW12,Long11}  
and also in a model of contextual quantum computing~\cite{W23_ur}. 

Using QCU also leads to certain issues.
A first nontrivial issue arises if the target quantum gate is unknown, i.e., a black box. 
This applies to situation of modular design, for instance. 
It was proven that quantum control over unknown gate is impossible~\cite{AFC14}. 
This is because the operation $U \mapsto \Lambda U$ is not valid
as it converts the unphysical global phase of $U$ to a local phase of $U$ in $\Lambda U$.
Here $\Lambda U$ is the controlled-$U$ gate.
A solution for this is to know an eigenstate of $U$,
which serves as a `flag' of it. 
The following circuit realizes the desired quantum control
\begin{figure}[h!]
    \centering
    \includegraphics[width=0.2\textwidth]{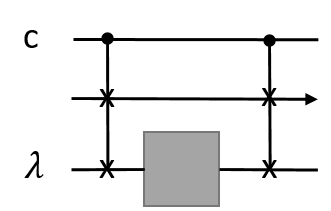}
\end{figure}

\noindent The top wire is the control register, the second wire is the data register,
and the third one is the flag.
Now the gate is `grey' instead of black since an eigenstate and also eigenvalue of it are known.

This method can be used to run quantum control over unknown programs, 
and then realize LCU algorithms. 
We require each program state $|U\ket$ is given with a flag $|\lambda_U\ket$. 
A flag state can be injected using our initialization method. 
See the figure for the linear combination of two unknown programs 
\begin{figure}[h!]
    \centering
    \includegraphics[width=0.3\textwidth]{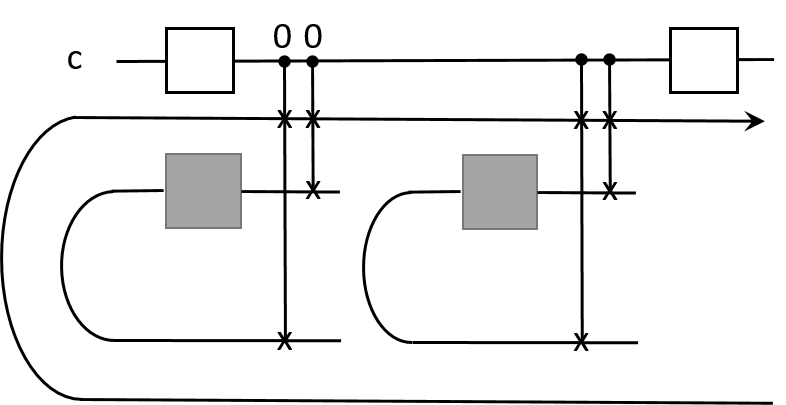}
\end{figure}

The control signal (c) itself is not pre-stored although it can be quantum.
That is, control signal is taken as deterministic input, 
and is not injected by measurement which is random.
Now the question is:
can it also be pre-stored as quantum states? 
For comparison, a random input data induced by PVM is okay 
since these input signals are orthogonal and their results are effectively equivalent. 
A random control signal induced by PVM will lead to uncertain operation,
say, $U_1$ or $U_2$, which are not orthogornal in general.
It seems there is no easy solution of this. 
In order to make orthogonal control signals, 
the nonlocal scheme~\cite{YRC20} can be used which turns a set of $\{U_i\}$  
into approximate orthogonal states,
but this requires a lot more quantum resources. 
Therefore, we do not require pre-stored quantum control signals.

For all the above, the control unit is required to be separable from the data unit at the output.
This is necessary since, by assumption, the control unit shall not carry final results.
They can get entangled during a computation, 
but at the end they shall be disentangled. 
This issue has been analyzed in the setting of quantum Turing machine~\cite{BV97,Mye97,Shi02,MW19,W20_Turing},
and also in a recent study of quantum control machine~\cite{YVC23}.
For instance, in the model of a local quantum Turing machine~\cite{W20_Turing},
by expressing the final quantum state carrying the results 
as a matrix-product state
\be |\psi\ket = \sum_i A(i_n)\cdots A(i_2)A(i_1)|\ell\ket |i_n\cdots i_2i_1\ket, \ee 
the machine register, with an edge state $|\ell\ket$ serving as the control register, 
can be disentangled from the data register at the end. 
Another method is to use measurement feedback to disentangle the controller and data~\cite{W23_ur} 
used to define a contextual quantum computing model. 
These examples also show that, due to entanglement, 
the interplay between the control flow and data flow 
needs more study.


\subsection{QPU}\label{subsec:cpu}

A CPU usually contains a control unit and ALU (arithmic logic unit).
For the classical case,
to run a program $A$, which is stored as bits $[A]$ in the hard disc,
the $[A]$ is firstly loaded as control and operations
on a programmable circuit, aka. chip.
The internal storage is also used to store temporary data. 
In this section, we study the primary structure of QPU in our model,
and compare with other existed ones. 

For the quantum case, 
the starting point is the circuit model. 
However, there are different approaches. 
We find there are two dual ones:
\begin{itemize}
\item The type-I: gates are stored as hardware while qubits are sort of `not there',
this applies to linear optics which uses optical elements as gates
and photons as qubits;
\item The type-II: qubits are stored as hardware while gates are sort of `not there',
this applies to SC, trapped ions which uses laser pulses 
(interacting with matter particles) as gates 
and particles (e.g., electrons) as qubits.
\end{itemize}

For both of them, a program $[U]$ that is used as control signals is classical.
We call this `classical programmability'. 
On the contrary, we will define a quantum programmability for our model of QPU.
It relies on the quantum encoding of programs, or a semi-quantum one,
namely, use Choi states $|H\ket$, $|T\ket$, $|CX\ket$ to store the elementary gates,
or other Choi states to store blocks of gates, 
and use bits $[U]$ to store their spacetime locations. 
Besides,
we need the toolboxes of switchable composition, quantum superchannel, and also quantum control unit.



We have seen that the control flow is distinct from the data flow. 
Actually, control sequences can also be stored as programs.
But in order to compose them, another level of controls are still needed
as long as the QPU is not automatic. 
That is, after all control signals are needed to monitor the evolution.
As discussed in the former section,
the input control signals need to be deterministic. 
That is, the $[U]$ is used as control signal to apply 
composition and other operations on primary Choi states.
To run $U$, qubits in the QPU will be measured.
After the run, qubits need to be refreshed to the right Choi states.

For instance, consider the programmable realization of a sequence of H and T gates 
to approximate a qubit rotation. 
See the figure 
\begin{figure}[h!]
    \centering
    \includegraphics[width=0.4\textwidth]{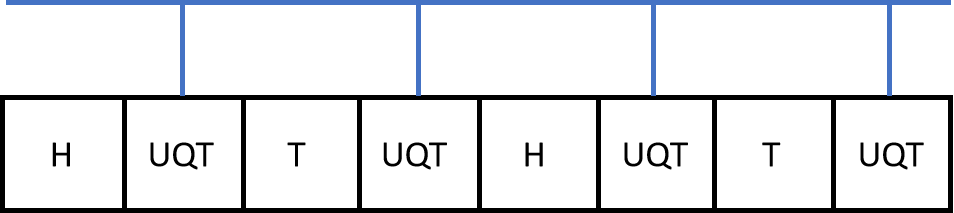}
\end{figure}

\noindent The UQT serving as the switchable composition is under control.
The elementary programs need to be refreshable in order to be multi-time programmable.
Furthermore, the QPU can also perform superchannels on the programs.
These can lead to quantum superalgorithms that prove to be powerful.
We will discuss this more in section~\ref{sec:alg}.

The `programmability' in our model is not classical. 
There are a few aspects for our programmability,
referring to the memory unit and control unit.
First, our model allows programs to be stored as quantum states, i.e., Choi states.
This means algorithms in our model can use quantum program states. 
Second, a stored quantum program is switchable, on or off,
depending on control signal. 
Third, the control signal can also be quantum,
so it forms an important part of a quantum algorithm. 

\begin{table}[b!]
    \centering
    \begin{tabular}{|c|c|c|}\hline
               & qubit & gate  \\ \hline 
    type-I     & time  & space \\ \hline
    type-II    & space & time  \\ \hline 
    type-III   & spacetime & spacetime  \\ \hline 
    \end{tabular}
    \caption{Comparison of three basic types of constructions of QPU.
    The type-III is used in QvN.}
    \label{tab:cpu}
\end{table}

Our construction of QPU lies somehow in between the type-I and type-II above.
We name this as type-III.
See the table~\ref{tab:cpu}.
For type-I, gates are stored as hardware, so they exist in the space domain,
while qubits such as photons are only generated on demand when an algorithm is to be run,
so they exist in the time domain.
It is the opposite for type-II. 
For the type-III, qubits are not used to carry data but are used to encode programs. 
The actual data qubits are prepared by measurements when an algorithm is to be run,
so they exist in the space and also time domain.
A gate also exists both in the form of a program and the operation in a composition,
so in the spacetime domain. 

What is the advantage of using type-III? 
At present, this is not fully clear yet. 
It apparently consumes a lot more qubits to realize an algorithm 
due to the usage of teleportation,
as the case of measurement-based model~\cite{RB01}. 
That said, teleportation shall bring some advantages (see section~\ref{subsec:tele}).
For instance, for the architecture design 
the physical qubits can tolerate more decoherence since they only carry 
the data for depth one before a composition occurs.
Also only qubits need to be manufactured, and gates are applied on them. 
Finally, there is a clear resource-theoretic characterization of QvN, 
by treating quantum memory as the universal resources~\cite{W23_ur}.
This can benefit the understanding of quantum superalgorithms 
which rely on quantum memory. 

\subsection{Program download/upload}

After a computation, the program state is consumed/destroyed.
This is not a flaw, however. 
This is also present in the usual quantum circuit model:
after the computation, the qubits need to be refreshed for the next task.
This even exists in classical computers which uses temporary data such as 
cache and buffer. 

For QvN, the program states are likely stored in the memory unit.
A computation would turn some of the states into garbage, basically.
The user has to restore the program.
This is achieved by downloading them through the internet from a software producer, or a host.
In order to be secure, here we consider quantum internet via quantum communication.
The usual carrier, although does not have to be, is photons.
Therefore, the user needs to have the ability to receive photons, store and measure them.
We find there are four primary schemes:
\begin{enumerate}
    \item use qubits to send bits: the host employs the bit-string description $[U]$ of a program $U$,
    and then encrypt the bits, $[[U]]$. 
    Quantum cryptography such as BB84 scheme can be used to send the bits~\cite{BB84}.
    At the user side, a control device is needed to receive the bits and apply the gate sequence,
    without revealing the bits to the user. 
    This is very much like a delegated computing or remote state preparation~\cite{GV19}, 
    but now the user is the remote site, and the host does not need to verify the user. 
     \item use ebits to send bits: one can use ebit-based quantum cryptography to send bit-string description of the gate sequence $[U]$. 
    \item use qubits to send qubits: the host prepare photons at the state $|U\ket$ directly,
    and send them to the user, who then applies quantum teleportation between photons and the memory qubits 
    to teleport/download the program from the photons to the memory qubits.
    \item use ebits to send qubits:
    the host and user first establish many pairs of ebits of photons,
    and then the user applies quantum teleportation between some photons and the memory qubits 
    to teleport/download the program from the photons to the memory qubits.
    Namely, if $|U\ket= V|0\ket$, the host applies $V^t$ and then the projection $|0\ket \bra 0|$ on his side,
    and that will prepare the photons at the user as $|U\ket$.
    The host needs to use our initial-state injection technique,
    and the effect on the final readout at the user's side can be easily dealt with.
\end{enumerate}

One may wonder which scheme is preferred.  
For the first two, the goal is to send bits,
which need to encode both the space and time information of the gate sequence in a program.
For the last two, the goal is to send qubits,
which does not need to encode the time information,
hence consuming fewer number of qubits than the number of bits apparently.
However, currently qubits are much more expensive than bits.
The choice of a scheme would depend on many practical conditions.

\begin{figure}[t!]
    \centering
    \includegraphics[width=0.4\textwidth]{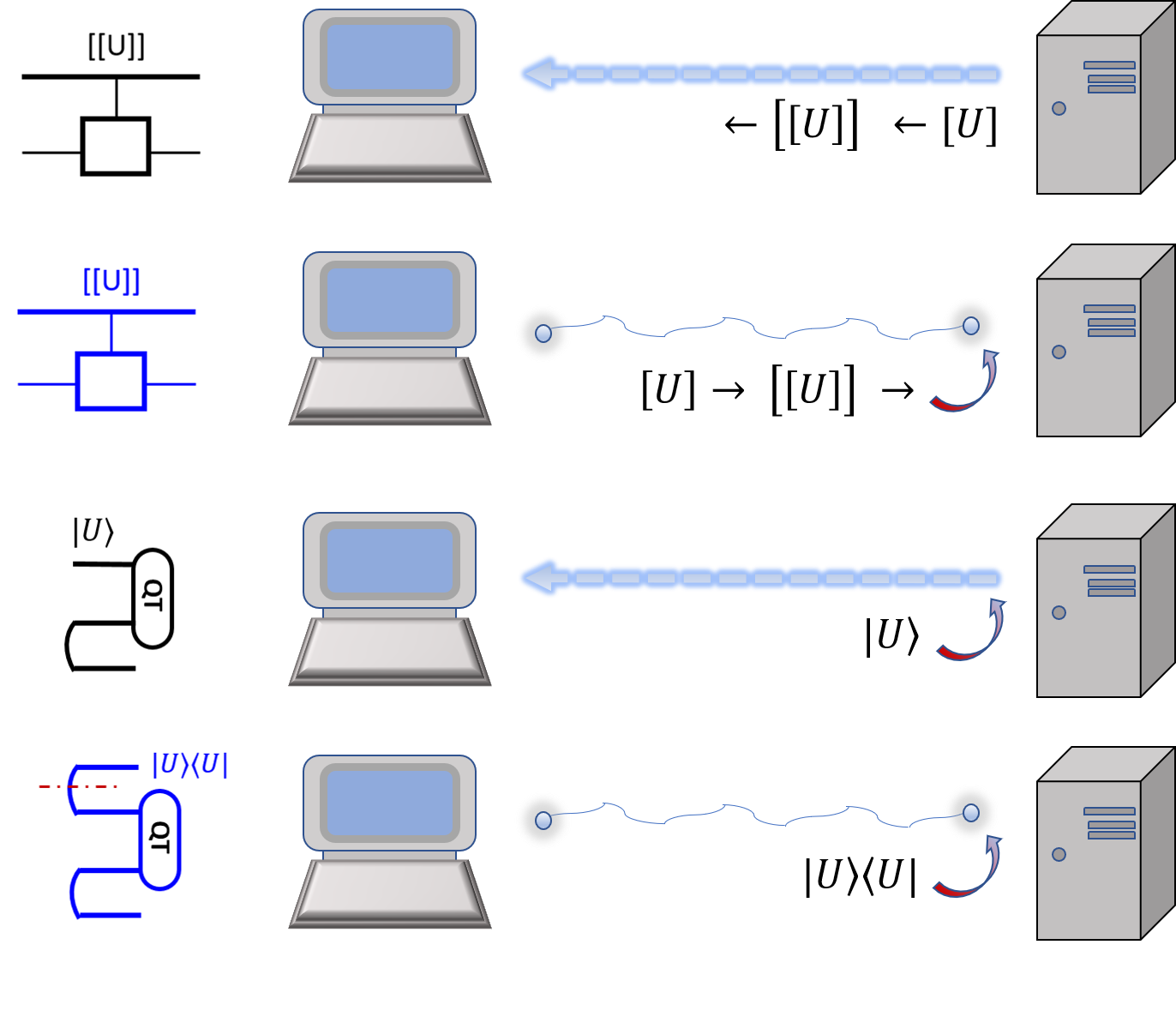}
    \caption{The four basic schemes to realize the download of quantum programs.
    The upload would be similar.}
\end{figure}


\subsection{Program verification}

When a user decides to download a program from a host,
the user has to verify that the host indeed has the promised program. 
This is a quantum verification task~\cite{GKK19,MSG+22},
which has been widely studied in recent years. 
Here we discuss how the program-verification would work,
but we do not specify all the details since 
there could be various schemes depending on practical settings. 

The verification can be interactive. 
In the framework of interactive-proof system~\cite{GKK19},
the user serves as a verifier,
and the host serves as a prover. 
Usually, the verifier is required to be computationally in BPP,
while the prover is in BQP. 
However, here in our setting the verifier is also in BQP 
but he only has a limited number of copies of the unknown program $|U\ket$.
That is, the user, as the verifier,
can only do verification instead of a full tomography. 

It is not hard to determine the number of samples of $|U\ket$ 
the user can download.
From verification theory which specifies an infidelity parameter $\epsilon$
and confidence parameter $\delta$,
the number of samples scales as  
\be N \in O\left(\frac{1}{\epsilon} \log \frac{1}{\delta}\right), \ee 
ignoring other factors that do not matter for our discussion here.
Although the scaling with respect to $\epsilon$ is not efficient,
for the purpose of verification a moderate fidelity is acceptable,
and the confidence is usually more important.

Given a few samples of $|U\ket$, the user can also do quantum estimation or learning. 
It is also well established that the fidelity scales as $N^{-2}$ 
for optimal joint global operations on them~\cite{CAPS04}.
This is the so-called Heisenberg limit. 

For a full process tomography, the user has to use a number of measurement operations,
hence a number of samples, 
that scales with the dimension of $|U\ket$,
which is exponential with the number of qubits it acts on.
Therefore, we find that 
as long as the number of samples is much smaller than 
that for tomography,
the verification can be done efficiently.
In addition, there is also another level of sampling 
which is to obtain the final expectation value of observable.
In modern terms this is a special instance of shadow tomography,
which can be done with a small number of samples~\cite{MSG+22}.

Verification is an important subject in the study of blind or delegated quantum computing.
We will study its difference from QvN in section~\ref{subsec:dqc}.

\section{Difference from other models}\label{sec:diffq}

In this section, we analyze the primary differences between QvN and some other models.

\subsection{Circuit model}\label{subsec:cir_mod}

Here we compare QvN with quantum circuit model (QCM), including the execution of an algorithm, security, 
verification, and other issues. 
We assume the usual scheme of QCM, 
which realizes a quantum algorithm as a three-stage process: initial state preparation, gate execution, and measurement.
We already see their difference from our study of programmable QPU. 

Actually, it is also fine to treat QvN in the framework of QCM,
as the primary operations are either unitary gates or measurements.
However, it is necessary to make a distinction between them since conceptually
QvN considered more requirements. 
This is similar for the classical case. 

The scheme to realize a circuit can be seen from this figure:
\begin{figure}[h!]
    \centering
    \includegraphics[width=0.2\textwidth]{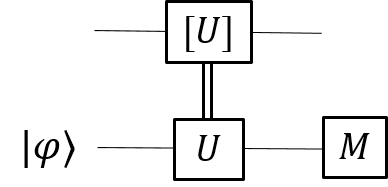}
\end{figure}

\noindent The top register is classical, $[U]$ is the classical representation of the program $U$. 
The output from a quantum algorithm is assumed to be the expectation value of a hermitian operator,
which reduces to the estimation of a set of probabilities, $p_i$.
The implementation of a quantum algorithm can be done efficiently provided that:
\begin{itemize}
    \item the initial state can be prepared quantum efficiently;
    \item the program $U$ can be stored classically efficiently, as $[U]$;
    \item the program $U$ can be realized quantum efficiently;
    \item the measurement for readout can be realized quantum efficiently;
    \item the number of samples scales efficiently to estimate each $p_i$.
\end{itemize}

The classical description $[U]$ is often a bit-string of the gate sequence 
in the circuit, if $U=\prod_i U_i$,
while bits are used to encode the spacetime location and type of each primary gate $U_i$
from a universal gate set, e.g., $\{$H,T,CZ$\}$.


After the execution, all qubits are measured and need to be refreshed for further usage.
The program is stored as bits $[U]$, so do not need refreshment, though. 
The composition of two programs $U_1$ and $U_2$ is simple, namely,
just implement them sequentially.
The initial state needs to be prepared before the application of gates,
which is not the case for QvN.

Another notable difference is that QvN requires the download of quantum programs,
since they cannot be cloned if bit-string description of them are unavailable. 
The security of quantum communication, relying on the uncertainty principle, 
ensures the security of the quantum programs. 
For the circuit model, 
a circuit is often not secure, 
i.e., there is a classical circuit \emph{diagram} which can be easily seen
and copied. 
However, there are also secure protocols relying on QCM. 
A notable example is the blind or delegated quantum computing (DQC)~\cite{BFK09},
which though was initially formulated via MBQC but can 
also be formulated via QCM. 
This leads to the discussion in the following two subsections. 





\subsection{MBQC}\label{subsec:mbqc} 

\begin{table}[b!]
    \centering
    \begin{tabular}{|c|c|c|c|c|}\hline 
             &  byproduct & type      & quantum program & switchability  \\ \hline 
    TBQC     &  Clifford  & 2-bit     & qubit gates     & no \\ \hline 
    MBQC     &  Pauli     & 1-bit     & CZ              & no \\ \hline 
    QvN      &  Pauli     & covariant & SU(d)           & yes \\ \hline
    \end{tabular}
    \caption{Comparison between QvN and MBQC.}
    \label{tab:mbqc}
\end{table}

Besides the circuit model, QvN also has close connections with MBQC. 
In the standard MBQC, also known as the one-way model~\cite{RB01},
a resource state such as the 2D cluster state is given,
and then a sequence of local adaptive measurements is performed to execute gates. 
The basic underlying mechanism is 1-bit teleportation~\cite{W19_rev},
and a spatial direction is chosen as the `teleported' evolution direction.
Furthermore, it is equivalent to the model based on Bell measurement for
the standard teleportation, which is 2-bit~\cite{CLN05}.
Here we will denote this as the teleportation-based model (TBQC),
although sometimes it is treated as a special case of MBQC.

For clarity, we summarized the comparison in Table~\ref{tab:mbqc}.
The TBQC is often used for fault-tolerant execution of gates,
hence its byproduct is extended to Clifford operations which still preserve Pauli gates.
Due to the covariant teleportation used in QvN,
the stored program can be fully quantum, namely, the whole group SU(d) of gates.
On the contrary, in MBQC qubit gates are induced by local measurements in rotated bases,
while in TBQC the entangling gates such as CZ or CNOT is done by two parallel Bell measurements 
consuming ebits. 
One shall note that in MBQC the local measurements are relatively simple, 
compared with Bell measurements, 
and the resource state can be prepared offine.

\begin{figure}[t!]
    \centering
    \includegraphics[width=0.7\textwidth]{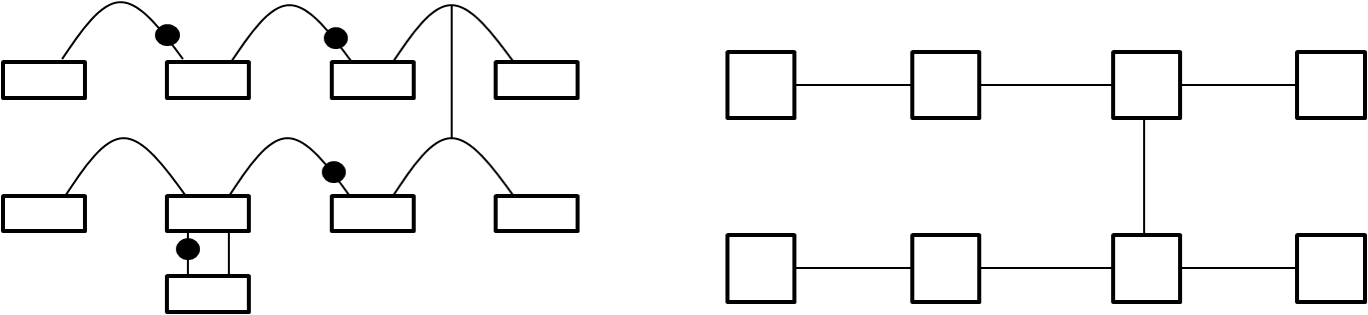}
    \caption{Information flow in QvN (left) and MBQC (right). 
    The vertical line represents the CZ gate.
    For QvN, a box represents a composition which is a measurement for teleportation.
    The curve with a dot represents a qubit program.
    A switchable program is also shown.
    For MBQC, a box represents a measurement on a site that realizes teleportation.
    }
    \label{fig:flow}
\end{figure}

Another important difference is that, 
in QvN the information, injected at the `tail', 
is always carried by the `head' of a Choi state.
There is no such explicit head-tail structure for MBQC, and also TBQC. 
The information flow is shown in Figure~\ref{fig:flow}. 
For QvN, the information never `cross' a composition `box',
but this is the opposite for MBQC. 
Treating a composition as a single depth, 
each physical qubit for a tail or head only has depth one. 
This leads to the switchability of the composition,
also illustrated in the figure. 
As has been discussed, the switchability could be useful 
to construct the QPU. 

\subsection{Delegated quantum computing}\label{subsec:dqc}

An important model for secure computation is the delegated quantum computing (DQC),
which was also much earlier known as blind quantum computing~\cite{BFK09}.
In this model, a user, as a verifier, aims to delegate computation to a prover 
without revealing the computation to the prover.
See the figure for QCM in subsection~\ref{subsec:cir_mod},
while the classical and quantum registers belongs to the verifier 
and prover, separately.
The verifier is in BPP while the prover is in BQP.
Usually, the verifier knows what to compute, 
but does not have the capability to do so. 
This model may apply to recent era of quantum computing 
that only a few labs or companies have powerful quantum computers,
and customers can use them blindly and confidently.
The input, output, and the computation itself can all be blind 
to the prover. 

This is different from the program-verification in QvN. 
For QvN, the host has both $[U]$ and $|U\ket$, 
while the user only has $|U\ket$. 
Th user will use the program blindly by making measurements on it.
Given limited samples of a program, the user can not do tomography,
i.e., cannot obtain its classical description $[U]$.
In DQC, the prover can do $U$,
which is equivalent to the ability to prepare $|U\ket$,
while the verifier has $[U]$.
In QvN, the user side is BQP, the host/prover is also BQP.  
The purpose of verification in DQC is to verify $[U]$,
while the purpose of verification in QvN is to verify $|U\ket$.
There is no apparent delegation in QvN.
See the table~\ref{tab:dqc}.

\begin{table}[b!]
    \centering
    \begin{tabular}{|c|c|c|}\hline 
            & user/verifier & host/prover \\ \hline 
     DQC     & $[U]$  & $|U\ket$ \\ \hline  
     QvN     & $|U\ket$  & $|U\ket$, $[U]$ \\ \hline 
    \end{tabular}
    \caption{Comparison between DQC and QvN.}
    \label{tab:dqc}
\end{table}

\section{Quantum algorithms in QvN}\label{sec:alg}

In this section, we study the design of quantum algorithms in QvN.
This has been analyzed in our previous work~\cite{W21_model,W22_qvn},
while here our discussion will be more specific,
drawing the connection with computational advantages. 

\subsection{Quantum superalgorithm}

A quantum algorithm is usually specified by a quantum circuit 
and a measurement procedure, as has been shown. 
On top of that, there is also a classical algorithm which designs 
the quantum algorithm. 
See the figure
\begin{figure}[h!]
    \centering
    \includegraphics[width=0.4\textwidth]{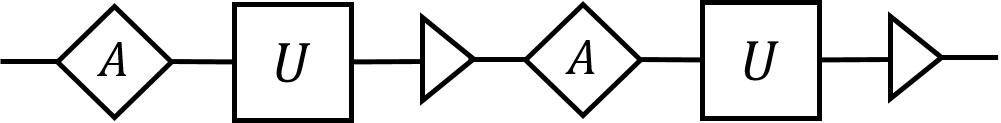}
\end{figure}

\noindent Here a triangle represents a measurement. This design can be iterative, with measurement outcomes feed forward 
to the classical algorithm, $A$,
which then optimizes the parameterized quantum circuit, $U$.
Some examples are the Solovay-Kitaev algorithm for gate compiling~\cite{DN06},
quantum channel simulation~\cite{WS15,WW23},
and quantum approximate optimization~\cite{FGG14}. 
This actually forms a classical comb of classical-quantum hybrid algorithm,
using the terminology of quantum superchannel theory~\cite{CDP08a,CDP08,CDP09}. 

One can also pose the following question:
can we use a quantum algorithm to design another quantum algorithm? 
Such a scheme works for the classical case, namely, 
there are classical algorithms that design classical algorithms. 
Such algorithms are often known as `meta' algorithms, or `hyper' algorithms,
since they contain some meta or hyper variables that need to be optimized. 
This plays essential roles in machine learning~\cite{MBW+19}.

For the quantum case, it has been confirmed that, indeed 
we can use a quantum algorithm to design another quantum algorithm. 
This follows from nothing but the quantum superchannel theory. 
The superchannel plays the role of the `meta' algorithm,
while the channels acted upon by the superchannel serve as the input to it.
We will call these algorithms as quantum superalgorithms 
to be consistent with the superchannel theory. 
It has the following structure 
\begin{figure}[h!]
    \centering
    \includegraphics[width=0.3\textwidth]{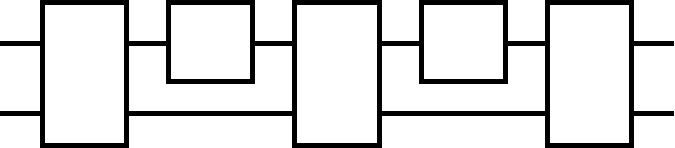}
\end{figure}

\noindent From the composition technique, 
it can also be realized by a sequence of composition~\cite{W22_qvn}. 
This is what a QPU can do, from section~\ref{subsec:cpu}. 
Compared with the former scheme above,
it is clear to see here quantum memory 
(the bottom register) are used as resources
to realize superalgorithms~\cite{W23_ur}.

Many quantum algorithms are of this form,
although sometimes they are not under a name of `superalgorithm'.
This includes quantum estimation and learning algorithms~\cite{MBW+19}, 
quantum channel discrimination~\cite{CAP08},
schemes for quantum games~\cite{GW07},
quantum optimization~\cite{LDR23}, 
quantum machine learning~\cite{DB18,VPB18,BLSF19,HKP21}. 
The recent quantum singular-value transformation~\cite{MRTC21} which can 
unify some quantum algorithms is also a special type of superalgorithm~\cite{W21_model}.

The theory of superchannel also allows the so-called higher-order operations~\cite{CDP09},
which are superchannels that act on superchannels,
by using the channel-state duality iteratively. 
These higher-order superchannel can still be viewed as superchannel
but with more complicated multipartite structure~\cite{W22_qvn}.

Besides, one may wonder that a `mother' algorithm is still needed 
to design a quantum superalgorithm,
and such a mother algorithm can be classical. 
This is indeed the case, but what matters is that 
quantum memory, and also control, are used as resources in addition 
to the current framework based on the circuit model. 
Recently, people find that quantum memory can lead to exponential advantages 
for solving some problems~\cite{HKP21,Car22}. 

Finally, the output for the final result often contains probabilities. 
This requests many runs of an algorithm to estimate them. 
However, there is a method to convert probabilities into amplitudes of quantum states, 
and then use the quantum amplitude estimation algorithm to obtain them
in the form of bit strings~\cite{BHM02}.
This is the analog of quantum phase estimation which can be used 
to estimate unknown parameters by encoding them as phase factors. 
Such algorithms use quantum controlled operations. 
From matrix decomposition, 
a controlled operation can be decomposed as the product of a few operations without control
and simple controlled operations such as the controlled-not gates. 
This means that these estimation algorithms can also be put into the framework of 
quantum superalgorithms. 



\subsection{Computational advantages}\label{sec:adv}

Finding computational advantages would depend on the structure of certain problems. 
If there is no structure, quantum computing can only provide quadratic advantages.
For instance, Grover's algorithm shows that for structure-less data base of size $N$,
a quantum computer will take time in order $O(\sqrt{N})$ instead of $\log N$~\cite{NC00}.
The so-called Heisenberg's limit from the uncertainty principle 
sets the bound for the precision of estimating an unknown parameter,
also with quadratic improvement of precision~\cite{TA14}. 
This relates to the fact that the square of amplitudes yields probabilities. 
In the model of QvN, 
we analyze the potential advantages by using the combination of 
quantum CPU, control, memory, and also internet, 
compared to the classical case and also other quantum computing models. 
Our analysis, however, is primary and we hope this can inspire more investigations.

\subsubsection{Storage}

It seems the amplitudes of qubits can significantly increase the ability for storage. 
A pure state $|\psi\ket=\sum_i \psi_i |i\ket$
already stored the amplitudes $\psi_i$ in it without encoding them as bit strings. 
However, this turns out not to be completely true,
since quantum measurements are needed in order to know $\psi_i$,
and this requires a lot copies of the qubit $|\psi\ket$.

The seminal result is the Holevo bound~\cite{Hol77},
which was established originally in the setting of quantum communication.
It states that the quantum capacity of a channel is half 
of its classical capacity. 
Namely, a qubit can only be used to store or transmit two bits. 
This may not be hard to understand from the point of view of error correction,
whereas correcting Pauli X and Z is enough to correct 
any linear combination of them.
This is also demonstrated by the quantum teleportation and dense coding~\cite{BBC+93}. 
Using an ebit, a qubit can be used to transmit two bits, 
and two bits can be used to transmit a qubit. 
It also relates to the quadratic speedup of Grover's algorithm. 
An $n$-qubit state can be viewed as $2n$ bits, 
hence representing $2^{2n}$ different values forming a database
whose size is the square of that from $n$ bits. 
Quantum search, if treated as a state prepration, 
cannot be faster since otherwise a qubit would carry more than two bits. 
However, this does not mean there is no larger advantages 
for specific tasks. 

In the circuit model, a state $|\psi\ket$ can be represented by its preparation circuit $U$.
As has been discussed, there is an efficient bit-string encoding $[U]$ of the circuit,
by only encoding the type of each gate and its spacetime location. 
However, using qubits to store quantum states can offer advantages, e.g.,
as we have mentioned for quantum learning~\cite{HKP21,Car22}. 
For instance, it has been proven although learning statistical average may not offer an advantage,
but accurately predicting the value of any observable can have an exponential advantage
by using quantum memory~\cite{HKP21}. 
Therefore, it is promising to explore more primacy of quantum memory. 

\subsubsection{Speed} 

Quantum advantage is often used the same as speedup.
This is basically because in principle 
other resource cost can be treated as a cost of computing time,
but still, we will distinguish them since this could inspire different intuitions for 
solving different problems.

First, we need to distinguish an algorithmic speedup from 
the speed of a quantum computer. 
The former is on the software level, 
while the later also depends on the hardware. 
The same quantum algorithm would take different amount of time 
on different quantum computers. 

The underlying mechanism for algorithmic speedup relates to interference. 
Compared to the classical case, 
quantum evolution is unitary, i.e., coherent, 
and there are significant amount of interference between trajectories, 
given a fixed basis of the underlying Hilbert space. 
A speedup occurs if the interference can enhance the probability 
of the desired trajectory.
Note that the pre-condition for a speedup is the accuracy of computing result.
The more accurate, the more time is needed. 

To achieve a speedup is one of the central tasks of QPU,
besides programmability. 
In the model of QvN, the interplay between QPU and other units 
will also affect the speedup.
This is also the case for classical computers,
and that is why cache is needed to replace the hard drive.
If a truly quantum computer can be built in the future,
following a certain QvN and the hierarchical design,
there will certainly be various kinds of quantum memory, 
control, communication, and even input and output devices, etc. 
The speed of such a quantum computer would depend on many factors 
that is still hard to sketch at present. 


\subsubsection{Security}\label{subsec:secure}

Quantum cryptography to realize security was 
one of the prompter of quantum information science,
with BB84 scheme of quantum key distribution as the notable example~\cite{BB84}.
It is based on the no-cloning theorem, 
which is equivalent to Heisenberg uncertainty principle. 

The program download/upload process can be seen as a special task in 
quantum communication.
So it shares features of standard quantum communication.
Here, we point out that it is secure in two senses.
First, the program generated by the host is secure against the user.
Second, the communication between the user and host is secure against Eavesdropper.
There are also other settings of secure quantum computing, 
such as delegated or blind quantum computing~\cite{BFK09,GMM13} discussed in section~\ref{subsec:dqc}
and multi-party quantum computing~\cite{CGS02}. 
For the later, neither one of the parties know the result.
Instead, they must communicate to extract the result.



Attacks in quantum cryptography have been well studied~\cite{GRT+02}.
An attack can be detected but not prevented. 
In the model of QvN, data are stored as qubits instead of bits.
Then one may be curious if quantum data can be hacked. 
Although qubits encoding data such as passwords cannot be accurately cloned hence leaked,
they can be measured. 
A simple quantum virus can be an instruction to make
the most trivial measurement which will erase any data. 
Such a virus is actually classical since it can take the same form and can copy itself.
Although there are costs to make an attack possibly at any time,
it cannot be prevented in principle. 
So we suppose that 
this stands as a `no-go' for a no-no-virus hope for quantum computers.
Despite this, quantum computers have potential for more applications in cryptography 
due to resources such as coherence and entanglement. 


\subsubsection{Energy} 

Reducing energy consumption in computation was one of the original motivation for 
quantum computing. 
Landauer showed that erasing bits will cost energy,
while with Toffoli gates, a classical computation can be made reversible~\cite{Ben80}.
It was at that time realized by Feynman and others 
that a quantum computer can be reversible since its evolution is unitary~\cite{Fey82}. 
However, compared with other features, 
the study of energy consumption in quantum computing is rare~\cite{TSS18,TSS20,CYR21}. 

In the circuit model, initialization and readout by measurement 
will cost energy.
A recent pioneer studied the energy cost in unitary evolution
using superchannel theory and resource theory~\cite{CYR21},
which showed that the energy cost relates to the accuracy of the computation. 
However, a systematic understanding of the thermodynamics of unitary evolution is lacking.
This is not obvious as thermodynamics often deals with non-unitary dissipative evolution.

Here, we point out that the energy issue could be relevant for the design of 
quantum control schemes, compared with classical ones.
However, it is not always straightforward to assess the amount of energy cost
since some schemes are `semi-classical.'
For instance, cooling a qubit by a reservoir in order to suppress decoherence
can be considered semi-classical. 
Using quantum error correction can also suppress decoherence,
but it is not easy to compare the energy costs during the cooling 
and the error correction.


In the model of QvN, quantum control unit is used to enact quantum operations,
and also to form ingredients of quantum algorithms.
When the control signal is not a part of the final output, 
erasing or resetting it will cost energy. 
On the contrary, using classical control cannot generate 
entanglement between the controller and the target. 
At present, it is not clear how to find quantum advantages on energy cost 
over classical control schemes.  

\section{NISQ implementation}\label{sec:nisq}

In this section, we study how to implement a small-scale QvN 
on noisy intermediate-scale quantum (NISQ) devices.
This would not include massive quantum error correction and quantum verification,
for instance, which require more quantum resources. 

We can compare to the basic requirements of the circuit model. 
This dates back to twenty years ago~\cite{Div00}, 
with five requirements:
a scalable system of qubits, 
initialization of qubits, 
sufficient coherence to carry out an algorithm,
universal set of unitary gates, and
measurement for read out.  

These requirements are also strengthened or expanded for more purposes. 
They are also the basic requirements to realize a QvN. 
A few additional requirements are needed. 
(1) First, it requires the ability to execute multi-qubit controlled gates,
such as the Toffoli gate. 
Such gates can be decomposed into elementary one or two-qubit gates,
but it would be better if they can be directly realized. 
These gates are needed for the initialization, composition, and also quantum control. 
(2) Second, it requires quantum communication.
This was also an extra request when flying qubits such as photons 
are needed to connect a few separate quantum stations,
such as in the trapped-ions setup~\cite{NC00}.
The above two requirements can already be satisfied by some systems~\cite{LKS+19,KM20,KMN+22}.
Therefore, it is possible to demonstrate prototypes of QvN.

Here we describe almost the smallest system of QvN,
with the process of read-write, download, composition, control, superchannel. 
They are listed as follows:
\begin{itemize}
    \item The read-write on program: 
            It needs two qubits to store a qubit gate, and four to store a CZ gate.
            The initial-state injection (i.e. write) for a qubit program on a standard basis ($|0\ket$ and $|1\ket$) 
            does not require ancilla, 
            and also the case for the read operation. 
            For the CZ program, the write operation on a standard basis ($|00\ket$, $|01\ket$, $|10\ket$, $|11\ket$) 
            does require a qubit ancilla and the Toffoli gate,
            but the read operation does not. 
    \item The download process:
            For the scheme using ebits to send qubits, 
            the state-injection at the host side requires Toffoli gate and an ancilla. 
            For a qubit program, the teleporation at the user side is on 4 qubits. 
            The download in total involves 9 qubits. 
            It is easy to verify for a two-qubit program, the teleporation is on 8 qubits. 
            The download in total involves 17 qubits. 
            For other schemes of the download, it requires less qubits hence also less gates.
    \item The composition: 
            To compose two qubit-program states deterministically,
            this needs 5 qubits with one as ancilla. 
            To compose a qubit program with the CZ program deterministically, 
            it needs 6 qubits if the qubit program applies earlier, 
            while 7 if it applies after the CZ. 
            However, if the Pauli byproduct is not required to be corrected, 
            less number of qubits are needed. 
            This reduces to 4 and 6, respectively. 
            In addition, to make the composition switchable, 
            extra ebits are needed. 
    \item  The quantum control: 
            To realize a quantum control of an unknown qubit program, 
            it requires 5 qubits, with one for a qubit control, 
            two for the qubit program, and two for the data register. 
            Recall that an eigenstate of the unknown qubit gate shall be known,
            and it will be injected by measurement. 
            For the control of an unknown two-qubit program, 
            it needs 9 qubits. 
    \item  The quantum superchannel:
            To realize an arbitrary qubit superchannel, 
            it needs 6 qubits, with two for the qubit program,
            and 4 as ancilla. 
            However, with a convex-sum decomposition algorithm~\cite{WW23},
            two ancillary qubits can be saved. 
    \item  A quantum superalgorithm: 
            For a simple quantum superalgorithm formed by a sequence of composition, 
            its cost is determined by the composition. 
            It can also include superchannels within such a superalgorithm,
            then its cost will be higher.
            For a simple demonstration, however, 
            the Pauli byproduct can be left uncorrected, 
            and even the initialization can be probabilistic. 
            This will realize a probabilistic or random superalgorithm.
\end{itemize}

We see that less than 20 qubits is enough to realize the primary operations in a QvN. 
Quantum systems nowadays already have far more qubits than this.
Therefore, more complicated operations, such as control or superalgorithm,
can also be realized. 



\section{Conclusion}\label{sec:conc}

To conclude, we presented a systematic survey of the recently 
introduced model of quantum von Neumann architecture. 
We put it in the more complete picture of a hierarchical design
principle of modern computers, 
which, given sufficient space and time, 
can not only realize universality, 
but also programmability, modularity, scalability, etc. 
We also briefly draw its connection with other 
quantum computing models and algorithmic advantages. 

On the theoretical side, there are also many interesting 
open questions. 
Here we list a few of them as our conclusion. 

\begin{itemize}
    \item Types of quantum memory unit. 
    A quantum RAM model of states was developed~\cite{GLM08},
    which could be faster to find a specific state than classical ones.
    Such a scheme can be used for the storage of Choi program states. 
    We mentioned there are various types of classical memory,
    and also memory devices. 
    This is not clear for the quantum case. 
    Our scheme for the quantum programs is more like the internal memory,
    instead of an external memory, i.e., a hard disc. 
    Although in the early days of computers, gates are indeed 
    applied on hard memory, 
    nowadays there is a clear distinction between internal 
    and external memory.
    It remains to investigate the role of external quantum memory. 
    \item The roles of quantum control. 
    As we have shown, using quantum instead of classical control unit 
    will cause issues such as the entanglement between the control flow
    and data flow. We also have mentioned a few tools to deal with this. 
    However, a general principle for the design of quantum control unit 
    is still needed. 
    Meanwhile, specific examples and application settings are also needed 
    to show the necessity of it, 
    instead of a classical one. 
    We pointed out energy consumption may relate to quantum control, 
    by studying the dynamics of work, heat, entropy, etc,
    i.e., the thermodynamics of quantum computing. 
    \item Quantum `sequential' circuit.
    A large class of classical circuits is known as the sequential circuits,
    which, roughly speaking, are circuits with memory or loop~\cite{HH13}. 
    They are essential for electric circuit design. 
    There is no apparent quantum analog as quantum circuits do not form loops, 
    despite some explorations~\cite{WLY22}. 
    Namely, an output from a quantum process cannot be an input again unless 
    it is trivial, i.e., it is a fixed point of the process. 
    This relates to the quantum closed timelike curve~\cite{Deu91}. 
    However, using Bell states and Bell measurements, 
    loops can be formed~\cite{W22_qvn}, as we have seen a Bell state or ebit
    is expressed as half of a loop. 
    The tricky part is that there are Pauli byproducts in Bell measurements. 
    Also using measurements makes the process non-unitary. 
    At present, it is unclear what could be the proper quantum notion of loop,
    leading to a quantum analog of classical sequential circuits. 
\end{itemize}

\section*{Acknowledgements}

This work is funded by
the National Natural Science Foundation of China under Grants
12047503 and 12105343.

\bibliography{ext}{}
\bibliographystyle{ieeetr}

\end{document}